\input harvmac
\Title{\vbox{\baselineskip12pt
\hfill{\vbox{
\vbox{\hbox{\hfil EFI-96-02}\hbox{hep-th/9601179}}}}}}
{\vbox{\hbox{\centerline{Chern-Simons from}}
\hbox{\centerline{Dirichlet 2-brane Instantons}}}}
\baselineskip 15pt plus 1pt minus 1pt
\centerline{Martin O'Loughlin}
\smallskip
\centerline{Enrico Fermi Institute}
\centerline{University of Chicago}
\centerline{5640 S. Ellis Ave.}
\centerline{Chicago, IL 60637 USA}
\centerline{\tt mjol@yukawa.uchicago.edu}
\bigskip
\baselineskip 12pt plus 1pt minus 1pt
\bigskip

In the vicinity of points in Calabi-Yau moduli space
where there are degenerating three-cycles the low
energy effective action of type IIA string theory
will contain significant 
contributions arising from membrane instantons
that wrap around these three-cycles. We show that the
world-volume description of these instantons is
Chern-Simons theory.
\Date{}
\vfill
\eject

\baselineskip 15pt plus 1pt minus 1pt
\newsec{Introduction}

Recent progress has led to a deeper understanding of the 
conifold transition in the type IIB superstring theory
\ref\conifold{A. Strominger, ``Massless Black Holes and 
Conifolds in String Theory'', {\it Nucl. Phys.} {\bf B451} (1995),
96, hep-th/9504090. B. Greene, D. Morrison and A. Strominger,
``Black Hole Condensation and the Unification of String Vacua'',
{\it Nucl. Phys.} {\bf 451} (1995), 109, hep-th/9504145.}. The 
conifold is a point in Calabi-Yau moduli space where an $S^3$
degenerates. One can pass through the conifold to the 
other side through a small resolution by blowing up an
$S^2$. In the mirror of this picture the conifold maps to 
a small resolution in IIA string theory. The image under
mirror symmetry of the 3-branes wrapping the $S^3$ providing
additional non-perurbative massless states is the 2-branes of the
IIA theory wrapping the $S^2$'s again giving additional massless
states.

\lref\bbs{K. Becker, M. Becker and A. Strominger, ``Fivebranes, 
Membranes and Nonperturbative String 
Theory'', {\it Nucl. Phys.} {\bf B456}
(1995), 130, hep-th/9507158.}
\lref\poldbrr{J. Polchinski, ``Dirichlet Branes
and Ramond-Ramond Charges'' {\it Phys. Rev. Lett.} 
{\bf 75} (1995), 4724, hep-th/9510017.}
\lref\wittbs{E. Witten, ``Bound 
States of Strings and p-branes'', hep-th/9510135}\lref\bsvI{M.
Bershadsky, V. Sadov and C. Vafa, ``D Strings on D Manifolds'',
hep-th/9510225.}\lref\bsvII{M. Bershadsky, V. Sadov and 
C. Vafa, ``D-branes and Topological Field Theories'', 
hep-th/9511222.}
On the ``other side'' of the conifold in the type IIA theory we 
have a 3-cycle blowing up, however we have no 3-branes in IIA
only even-branes. Therefore the important non-perturbative object 
to consider on this side of the conifold is the instanton
formed by wrapping the Dirichlet membrane (D2-brane)   
world-volume of the IIA string theory
\poldbrr\ around the appropriate 3-cycle. We propose to study these
D2-brane instantons by introducing their description via 
Chern-Simons theory. Other work on membrane instantons may be
found in \bbs\ though in that work the techniques used are
significantly different to those we propose here. We believe that
our methods will be complimentary to theirs and also to the 
other methods recently developed for the study of Dirichlet 
branes \refs{\wittbs, \bsvI, \bsvII}.

How do we describe strings attached to a Dirichlet 2-brane instanton 
that is wrapped around a 3-cycle of a Calabi-Yau manifold?
It was shown by Polchinski \poldbrr\ that the cycle around
which the Dirichlet-brane is wrapped must be supersymmetric and also that
the D-brane carries Ramond-Ramond charge.
We will consider D2-brane instantons and will look 
for an open string field theory that describes the interaction 
between  the strings and this 
supersymmetric field configuration on the boundary.

\lref\wittopen{E. Witten, ``Open Strings as Chern-Simons'', 
hep-th/9207094.}
There is an open string theory that specifically fulfils the
requirement that the string endpoints rest on a supersymmetric 
3-cycle (actually a Lagrangian three-fold this being a necessary 
condition for the submanifold to be a supersymmetric 3-cycle) 
of the Calabi-Yau manifold. It was obtained
by Witten \wittopen\ by
topologically twisting the bosonic open string theory by an 
A-type twisting. One can show that this open string theory has 
a corresponding open string field theory that is simply Chern-Simons 
theory on the 3-cycle. We show that the 
instanton corrections to the effective field theory, in particular
near points in moduli space where there are degenerating 3-cycles, 
may be calculated using Chern-Simons theory on the appropriate 
3-cycle. 

\newsec{The Details}

What are the properties of the three cycle around which we are wrapping 
the membrane? The three cycles that 
appear are those for which the wrapping of a D2-brane
will not break more than half the supersymmetries of the string theory
\lref\jjj{J. Hughes, J. Liu and J. Polchinski, ``Supermembranes'',
{\it Phys. Lett.} {\bf B180} (1986), 370.}
\refs{\jjj, \bbs, \bsvII}. This condition is the same as 
requiring that some global supersymmetry transformation
restricted to the 3-cycle can be undone by a $\kappa$-transformation.
Let $i:M\rightarrow X$ be an embedding of a 3-cycle $M$ into a 
Calabi-Yau manifold $X$. In this case the condition of a 
supersymmetric three cycle is the requirement that
$i^* \Omega \sim V$ and
$i^* \omega = 0$, where $V$ is the volume of the three cycle, 
$\Omega$ is the holomorphic three form on $X$
and $\omega$ is the Kahler form on $X$. Such cycles minimize the 
p-brane world-volume \bbs. For a 3-cycle to be a Lagrangian 
submanifold we need $i^* \omega = 0$ and thus supersymmetric
3-cycles are three-dimensional Lagrangian submanifolds. 

For the A-twisted bosonic string with Dirichlet boundary conditions in M,
BRS invariance of the boundary conditions is fulfilled if the three
manifold described by the boundary conditions is a
Lagrangian submanifold with a flat connection living on it. 
This was explained by Witten in \wittopen\ and
we  will explain it in some detail 
below. An earlier construction where the 3-manifold is found as 
the fixed set of an involution is described in 
\ref\horava{P. Ho\v{r}ava, ``Equivariant Topological Sigma 
Models'', {\it Nucl. Phys.} {\bf B418} (1994), 571, originally 
PRA-HEP-90/18, hep-th/9309124.}\ in which case the 3-manifold 
was a Lagrangian submanifold by construction.
Thus we have discovered that the A-twisted open string 
is the appropriate description for strings in the presence of 
D2-brane instantons wrapped around a supersymmetric 3-cycle.
As discussed in \ref\bcov{M. Bershadsky, S. Cecotti, H. Ooguri and
C. Vafa, ``Kodaira-Spencer Theory of Gravity and Exact Results for
Quantum String Amplitudes'', {\it Comm. Math. Phys.} {\bf 165} (1994),
311, hep-th/9309140}\ the topologically
twisted string theory calculates certain quantities in the 
supersymmetric string theories. For the A-type twisting corrections
to the effective action of the type IIA superstring are obtained. 

We will now follow the discussion of Witten in \wittopen\
and describe in detail 
the boundary theory of the A-twisted string theory with 
a Calabi-Yau target space. 
As above, let $X$ be the Calabi-Yau target space and
$\Phi : \Sigma\rightarrow X$ the map from the Riemann 
surface into the target space. In the A-twisted 
string theory we have the three fields $\phi$ (the local
components of $\Phi$), $\chi$ and $\psi$
with the following fermionic symmetry ($I, J$ are co-ordinate indices
on $X$, $x^\alpha, z, {\bar{z}}$ co-ordinate indices  on $\Sigma$ and 
$i, {\bar{i}}$ spinor indices on $X$), 
\eqn\qtrans{\eqalign{
\delta \phi^I &= i\alpha\chi^I \cr
\delta \chi^I &= 0 \cr
\delta \psi_z^{\bar{i}}  &= -\alpha \del_z \phi^{\bar{i}} - i\alpha 
\chi^{\bar{j}} \Gamma_{\bar{j}\bar{m}}^{\bar{i}} \psi_z^{\bar{m}}\cr
\delta \psi_{\bar{z}}^i  &= -\alpha \del_{\bar{z}} \phi^i - i\alpha 
\chi^j \Gamma_{jm}^i \psi_{\bar{z}}^m\cr
}}
The boundary condition is $\partial\Sigma \rightarrow M$ where $M$ is 
required to be a Lagrangian submanifold. 
We define $Q$ as the generator of these transformations; $\delta\Lambda 
= -i\alpha \{Q,\Lambda\}$.
Our Lagrangian is given by $L = i\{Q,V\}$ and $T_{\alpha\beta} = 
i\{Q,b_{\alpha\beta}\}$, where $V = t\int_\Sigma d^2z g_{i\bar{j}}
(\psi_z^{\bar{i}} \del_{\bar{z}}\phi^j + 
\del_z\phi^{\bar{i}}\psi_{\bar{z}}^j)$ and
$b_{\alpha\beta} =
itg_{IJ}(\psi_\alpha^I\del_\beta\phi^J + \psi_\beta^I\del_\alpha\phi^J
- h_{\alpha\beta}h^{\sigma\tau}\psi_\sigma^I\del_\tau\phi^J).$

In the 3-manifold M there is a gauge field coming from the 
Chan-Paton factors of the open string theory. Denote this field 
by $A$ and its field strength by $F$. From the recent 
study of p-brane bound states \wittbs\
we expect an n-wrapped D2-brane to have associated with it
an open string theory with a $U(n)$ Chan-Paton factor. 
This non-abelian group arises from the n different leaves of the
multiply wrapped 2-brane to which the open string may attach. 
For the soliton world volume theory formed by pulling back the target
space effective action to the world-volume of a Dn-brane wrapped about
an n-cycle \wittbs\ the
multiply wrapped configuration was not allowed because it was argued 
to not produce the correct counting of BPS states and to have an 
inconsistent world-volume field theory. For instantons we claim that 
to be saddle points the membranes must wrap Euclidean supersymmetric 
3-cycles and as stability is not an issue for instantons, multiple 
wrappings of the membrane world-volume around the supersymmetric
3-cycle are allowed. 

The boundary term in this twisted open string theory 
is a Wilson line running around the boundary
of $\Sigma$ and the variation of the boundary term under the 
fermionic symmetry is
\eqn\boundvar{
\delta(Tr P exp \oint_{\partial\Sigma} \Phi^*(A)) = 
Tr(\oint_{\partial\Sigma} d\tau\,  
i\alpha\chi^I {d\phi^J\over d\sigma} F_{IJ}(\tau) 
P exp \oint_{\partial\Sigma; \tau} \Phi^*(A)),
}
and $\oint_{\partial\Sigma; \tau}$ means the integral around 
$\partial\Sigma$ starting and ending at $\tau$. 
The BRS invariance of this boundary condition shows that the 
world-sheet theory can be consistently
coupled to flat connections on the D-instanton world-volume.

We will describe the corresponding string field theory. To do this 
we first look at the eigenfunctions of $L_0$,
\eqn\virop{
L_0 = \half \int_0^\pi (-{1\over t} g^{IJ} {\delta^2\over
\delta\phi^I\delta\phi^J} + t g_{IJ} {d\phi^I\over d\sigma}
{d\phi^J\over d\sigma} + \dots ).
}

For weak coupling ($t\rightarrow\infty$) where the theory 
is exact $L_0 = 0$ requires that
${d\phi^I\over d\sigma}=0$ and the entire world-sheet is 
mapped into the 3-cycle. 
We can then write the string field functional 
in the three-cycle as, 
${\cal A} = c(q) + \chi^a A_a(q) + \chi^a \chi^b B_{ab}(q) + \dots $ 
where $q$ is a local co-ordinate on the three manifold $M$.
With the Chan-Paton factors for the case where we have the 
D2-brane wrapped multiple times around the three-cycle $A_a$
is a non-abelian gauge field on $M$. Also $\{Q,\phi\} = -\chi$ and
$\{Q,\chi\} = 0$ which are solved on $M$ by $Q=d$ and $\chi = -d\phi$.
With these
relations plugged into the open string field theory action
the open string world lines in $M$ look just like the 
fattened matrix propagators of the matrix field $A$ in $M$,
\eqn\osf{\eqalign{
S_{osft} &= {k\over 2} \int ( {\cal A}\star Q{\cal A} + {2\over 3} 
{\cal A}\star{\cal A}\star{\cal A}) \cr &=
{k\over 2} \int_M (A\wedge dA + {2\over 3} A\wedge A\wedge A)}}

The total membrane contribution will then involve the 
Chern-Simons theory plus the membrane self-energy coming from the 2+1 
supermembrane action, analytically continued to the Euclidean metric,
\eqn\Smem{
S_{mem} = {T_{mem}\over 2}\int d^3\sigma \sqrt{\gamma} 
(\gamma_{ij} \del_m X^i
\del_n X^j - \half + \gamma^{ij}\gamma^{kl}F_{ik}F_{jl}) + \dots}
Wrapping the membrane world-volume around the 
Lagrangian three-cycle sets $F=0$ (from \boundvar\ or \osf) 
and thus the membrane self-energy
weighting is the action of the three-cycle world-volume. 
For a generic point in the moduli space 
of the Calabi-Yau manifold, we may expect that the world-volume
instantons are not dominant in the partition function due to the weight
of the world-volume coming from \Smem. 
By the Lagrangian submanifold condition the three-cycle has 
minimal volume and the instanton corrections can therefore
become significant near points
in the moduli space where a three-cycle is degenerate. 
We claim that these corrections can be 
obtained from the Chern-Simons theory
described above, with an additional contribution coming from the
supermembrane world-volume theory. 
\lref\poldbounds{J. Polchinski, ``Combinatorics of Boundaries in 
String Theory'', {\it Phys. Rev.} {\bf D50} (1994), 6041, 
hep-th/9407031.}

Putting all these pieces together and using the dilute instanton
gas approximation to carry out the instanton summation, 
the total expression for the 
D2-brane instanton correction to the partition function 
when the instanton is obtained by wrapping a 3-sphere is
\eqn\totcorr{
Z_{inst} = \sum_{m=1}^\infty (KV)^m \sum_{\{n_i\}_1^m} 
\prod_{j=1}^m e^{- n_j S_{mem}} Z[U(n_j),k;S^3] \sqrt{n_j S_{mem}}} 
where
\eqn\ZnkSSS{
Z[U(n),k;S^3] = (k + n)^{-n/2} \sqrt{(k + n)\over n} \prod_{j=1}^{n-1} 
(2 sin({j\pi \over n+k}))^{n-j},}
$V$ is the volume of the space-time in which the instantons live, K is 
a factor that arises from the closed string fluctutations around the 
instanton configuration
and $S_{mem}$ is evaluated for a D2-brane wrapping the
minimal 3-cycle of a Calabi-Yau manifold (we are 
considering for simplicity the case where we have $h_{2,1}=1$). We 
assert that $K$ is a constant independent of the wrapping number. In 
field theory $K$ comes from the correction to the flat space 
field fluctuations caused by the presence of the localized 
instanton configurations. As the instanton size is fixed by the string
gas that surrounds it, and as this is not strongly dependent upon
the number of times the membrane wraps the 3-cycle, we expect $K$
to be constant. The Chern-Simons 
coupling constant $1/\sqrt{k}$ is equal to
the open string coupling constant which is in turn equal to the 
square root of the closed string coupling. The D-brane tension
$T_{mem}$  which 
appears in the membrane action \Smem\ is 
proportional to the disk amplitude of the string theory 
(with the disk boundary resting on the Ramond-Ramond charged D-brane)
as argued by Polchinski \poldbounds, 
which is in turn proportional to $1/g_{st}$. So the 
$e^{-n S_{mem}}$ weighting factor with
$S_{mem} \sim {1\over g_{st}} V_{3-cycle}$ 
is the source of the $e^{-1/g_{st}}$ corrections argued to be 
present in string theory by Shenker \ref\ssmm{S. Shenker, ``The Strength
of Nonperturbative Effects in String Theory'', In {\it Cargese 1990, 
Proceedings, Random surfaces and quantum gravity}, 191.}.
This will contribute significantly to the 
partition function when the volume of the compactification 
3-cycle is small,
for example for the type IIA string near the conifold point in 
the moduli space of the quintic threefold. 

The instantons that arise bear a passing resemblance to the 
instantons of the Fermi sea picture of the $c=1$ string theory. This
similarity is somewhat strengthened by recent conjectures relating
the $c=1$ model at the self-dual radius to the string 
physics of the conifold transition. 
The instantons of $c=1$ arise when one continues the Fermi 
sea to Euclidean space; $uv = \mu$ becomes $x^2 + y^2 = \mu$. 
The instanton is a single fermion that runs around the circular
Fermi sea of this Euclidean continuation. For the conifold in the 
IIA theory we have a 3-sphere $\sum_{i=0}^3 x_i^2 = \mu$ about
which our membrane instanton is wrapped, and this 3-sphere is 
conjectured \ref\gv{D. Ghoshal and C. Vafa, ``$c=1$ String 
as the Topological Theory of the Conifold'', {\it Nucl. Phys.}
{\bf B453} (1995), 121, hep-th/9506122.}\ to be in a direct 
relationship to the Fermi sea of $c=1$
and also to the ground ring \ref\wittgr{E. Witten, ``Ground 
Ring of Two-dimensional String Theory'', {\it Nucl. Phys.} {\bf B373}
(1992), 187, hep-th/9108004.}\
of $c=1$. This implies an
identification between $k$ and $\mu$. With this in hand the 
conifold is approached by taking $\mu\rightarrow 0$ 
and thus $k\rightarrow 0$. In this
limit the partition function \ZnkSSS\  simplifies somewhat when $n$
is also large
\ref\periwal{V. Periwal, ``Topological Closed String Interpretation
of Chern-Simons Theory'', {\it Phys. Rev. Lett.} {\bf 71} (1993),
1295, hep-th/9305115.}\ 
and the ubiquitous Bernoulli numbers arise, giving hope that some 
closed form expression may be obtained\foot{A relationship between 
Chern-Simons theory, the conifold and the $c=1$ string at the 
self-dual radius was also observed recently in \ref\djbp{D. Jatkar
and B. Peeters, ``String Theory near a Conifold singularity'', 
{\it Phys. Lett.} {\bf B362} (1995), 73, hep-th/9508044.}. This 
proposal is distinct from ours in that ours involves the ``other
side'' of the conifold.}.

The dilute instanton gas approximation that we have used to 
derive the expression for $Z_{inst}$ should be very good.
The main requirement for the absence of 
instanton interactions is that the size of the instantons be ``small''.
In our case the instanton is wrapped around a cycle of the internal 
Calabi-Yau manifold and so is essentially pointlike from the point
of view of the four-dimensional physics. Even taking account of the
strings attached to the instanton its size will be no 
greater than the string scale and thus one would expect the dilute 
instanton gas approximation to be excellent.

\newsec{Further Directions}

For more general degenerations of the type discussed in 
the recent literature  one needs to consider in addition 
to the open strings attached to a single Dirichlet instanton
the strings with different ends attached to different 
instantons. Obviously when the individual D-instantons are widely
separated such contributions will be supressed by the world-volume
of the stretched string. Close to enhanced symmetry 
points and moduli space singularities these contributions will be 
important. In addition to the
weighting factor that comes from the supermembrane world-volume 
theory, we must allow the string field theory that lives on the 
instanton world-volume to include diagrams where only one end of the 
string rests in a particular manifold. Thus we need to add to the 
Chern-Simons theory some particles living in the fundamental 
representation of the gauge group. We also need to consider
wrapping the membranes around non-trivial 3-cycles with 
topology other than the $S^3$ considered here. 
In \bsvI\ some of the 3-cycles have the 
topology of $S^2 \times S^1$. The partition function of \ZnkSSS\ is then 
modified \ref\jonespoly{E. Witten, ``Quantum Field Theory and the Jones 
Polynomial'', {\it Comm. Math. Phys.} {\bf 121} (1989), 351.}.
According to the discussion of \wittopen\ in situations where the 
Lagrangian 3-fold has non-trivial topology, as for $S^2 \times S^1$,
additional topological terms can be added to the string field theory
action. The significance of such additional terms needs to be understood 
in this context. 

\lref\mrd{M. Douglas, ``Branes within Branes'', 
hep-th/9512077.}
In the spirit of using field theory on world-volumes to discuss 
string theory interacting with D-branes \refs{\wittbs, \bsvI, 
\bsvII, \mrd}
we may try to extend the 
Chern-Simons theory for instance by coupling it to some 
fermions living on the world-volume. With additional matter present we 
have the possibility of phase transitions and we may 
inquire as to whether these have some 
interpretation in terms of D-brane bound states or collective
excitations of D-branes.

Within the Chern-Simons theory 
we can also calculate the contributions of
these four-dimensional instanton (point events in the four-dimensional
space-time) configurations to various coupling constants of
the low energy effective action (for example the calculation of 
the instanton correction to the four-fermi interaction in the 
low energy effective theory of IIA strings in four dimensions in 
\bbs). 
Such instanton corrections to string scattering should be found by 
calculating Wilson loop correlators in the Chern-Simons theory 
\jonespoly. In more generality we could investigate the meaning of 
the Wilson loop correlators that one calculates in the Chern--Simons 
theory. The skein relations of the Chern-Simons theory allow
us to unknot configurations of Wilson loops introducing
M\"obius band configurations. For the two loop correlators this seems
to indicate that the linked loops calculate the full answer and 
the unlinked loops calculate the correction excluding 
the unoriented world-sheets attached to the D2-brane 
instanton. Certainly the partition
sum includes these unoriented configurations. 

Finally we need a means to investigate the small structure 
of the type conjectured 
recently in \ref\hardconj{S. Shenker, 
``Another Length Scale in String Theory?'', hep-th/9509132.}\ or as
observed within the context of Calabi--Yau moduli space in 
\ref\agmsml{P. Aspinwall, B. Greene and D. Morrison, ``Measuring Small 
Distances in $N=2$ sigma models'', 
{\it Nucl. Phys.} {\bf B420} (1994), 184, hep-th/9311042.}. 
Our instantons wrapped about the collapsing three--cycles 
will become $-1$--branes in the limit of zero volume. 
Such configurations are conjectured in \hardconj\ to give rise to 
point--like structure in string theory. By appropriate manipulation 
of the Chern--Simons theory we may develop some tools to 
investigate this conjecture. Further development of the 
relationship between the D2-brane instantons and the 
eigenvalue instantons of the $c=1$ matrix model is also desirable. 
There is certainly much more to be done than has been done.

\centerline{\bf Acknowledgements}

I would like to thank Jeff Harvey, Albion Lawrence and 
especially Emil Martinec for discussions. This work was supported 
by DOE grant DE-FG02-90ER40560.

\listrefs

\end